\documentclass[runningheads]{cl2emult}

\makeindex            % used for the subject index
\def\go{\mathrel{\raise.3ex\hbox{$>$}\mkern-14mu\lower0.6ex\hbox{$\sim$}}}
\def\lo{\mathrel{\raise.3ex\hbox{$<$}\mkern-14mu\lower0.6ex\hbox{$\sim$}}}

\begin{document}
\title*{The Final Fate of Coalescing Binary Neutron Stars:
Collapse to a Black Hole?}
\toctitle{The Final Fate of Coalescing Binary Neutron Stars}
% allows explicit linebreak for the table of content
%
%
\titlerunning{Coalescing Binary Neutron Stars}
% allows abbreviation of title, if the full title is too long
% to fit in the running head
%
\author{Frederic A.~Rasio}
\authorrunning{Frederic A.~Rasio}
% if there are more than two authors,
% please abbreviate author list for running head
%
%
\institute{Department of Physics,
Massachusetts Institute of Technology, Cambridge, MA 02139, USA}

\maketitle              % typesets the title of the contribution

\begin{abstract}
Coalescing  compact 
binaries with neutron star (NS) or black hole (BH) components
are important sources of gravitational waves for the
laser-interferometer detectors currently under construction, and may
also be sources of gamma-ray bursts at cosmological distances.
This paper focuses on the final hydrodynamic coalescence and
merger of NS--NS binaries, and
addresses the question of whether black hole formation is
the inevitable final fate of these systems.
\end{abstract}

\section{Introduction}

Many theoretical models of gamma-ray bursts (GRBs) 
rely on coalescing compact 
binaries (NS--NS or BH--NS) to provide the energy of GRBs at
cosmological distances (e.g., Eichler et al.\ 1989; Narayan,
Paczy\'nski, \& Piran 1992; M\'esz\'aros \& Rees 1992). 
The close spatial association of some GRB afterglows 
with faint galaxies at high redshifts is not inconsistent
with a compact binary origin, in spite of the large recoil
velocities acquired by compact binaries at birth (Bloom,
Sigursson, \& Pols 1999).
Currently the most popular models all assume that the coalescence leads
to the formation of a rapidly rotating Kerr BH surrounded by a torus
of debris. 
Energy can then be extracted either from the rotation of the BH or from
the material in the torus so that, with sufficient beaming, the
gamma-ray fluxes observed from even the most distant GRBs can be
explained (M\'esz\'aros, Rees, \& Wijers 1999). However, it is important to
understand the hydrodynamic processes taking place during the final 
coalescence before making assumptions about its outcome. In particular,
as will be argued in \S 3, it is not clear that the coalescence of
two $1.4\,M_\odot$ NS will form an object that must collapse to a BH
on a dynamical time,
and it is not certain either that matter will be ejected
during the merger and form an outer torus around the central object.

Coalescing compact binaries are also 
the most promising sources of gravitational
waves for detection by the large laser interferometers 
currently under construction,
such as LIGO (Abramovici et al.\ 1992) and VIRGO (Bradaschia et al.\ 1990). 
In addition to providing a major new confirmation of
Einstein's theory of general relativity (GR), including the first direct
proof of the existence of black holes (Flanagan \& Hughes 1998;
Lipunov, Postnov, \& Prokhorov 1997), the detection of gravitational
waves from coalescing binaries at cosmological distances could provide 
accurate independent measurements of the Hubble constant
and mean density of the Universe (Schutz 1986; Chernoff \& Finn 1993; 
Markovi\'c 1993). 
Expected rates of NS--NS binary coalescence in the Universe, 
as well as expected event rates in laser interferometers, have 
now been calculated by many groups. Although there is some disparity 
between various published results, the estimated rates are generally 
encouraging (see Kalogera 2000 for a recent review).
Many calculations of gravitational wave emission from coalescing binaries 
have focused on the waveforms emitted during the last few thousand orbits, 
as the frequency sweeps upward from $\sim10\,$Hz to $\sim300\,$Hz.
The waveforms in this frequency range, where the sensitivity of
ground-based interferometers 
is highest, can be calculated very accurately by 
performing high-order post-Newtonian (PN)
expansions of the equations of 
motion for two {\it point masses\/} (see, e.g., Owen \& Sathyaprakash 1999
and references therein). However, at the end of the inspiral, 
when the binary separation becomes comparable 
to the stellar radii (and the frequency is $\go1\,$kHz), 
hydrodynamics becomes important and the character 
of the waveforms must change. 
Special purpose narrow-band detectors that can sweep up frequency in real 
time will be used to try to catch the last $\sim10$ cycles of the gravitational
waves during the final coalescence
(Meers 1988; Strain \& Meers 1991). These ``dual recycling''
techniques are being tested right now on the German-British interferometer
GEO 600 (Danzmann 1998). In this terminal phase of the coalescence,
when the two stars merge together into a single object, 
the waveforms contain information not just about the 
effects of GR, but also about the interior structure 
of a NS and the nuclear equation of state 
(EOS) at high density. 
Extracting this information from observed waveforms, 
however, requires detailed theoretical knowledge about all relevant
hydrodynamic processes. 
If the NS merger is followed by the formation 
of a BH, the corresponding gravitational radiation waveforms will also 
provide direct information on the dynamics of rotating core collapse
and the BH ``ringdown'' (see, e.g., Flanagan \& Hughes 1998).

\section{Hydrodynamics of Binary Coalescence}

The final hydrodynamic merger of two NS is driven by a combination
of relativistic and fluid effects. Even in Newtonian gravity,
an innermost stable circular orbit (ISCO) is imposed by
{\it global hydrodynamic instabilities\/}, which can drive 
a close binary system to rapid coalescence once the tidal interaction 
between the two stars becomes sufficiently strong.
The existence of these global instabilities 
for close binary equilibrium configurations containing a compressible fluid, 
and their particular importance for binary NS systems, 
were demonstrated for the first time by 
Rasio \& Shapiro (1992, 1994, 1995; hereafter RS1--3) 
using numerical hydrodynamic calculations.
These instabilities can also be studied using analytic methods.
The classical analytic work for close binaries containing an
incompressible fluid (e.g., Chandrasekhar 1969) was
extended to compressible fluids in the work of Lai, Rasio, \& Shapiro 
(1993a,b, 1994a,b,c, hereafter LRS1--5).
This analytic study confirmed the existence of dynamical 
instabilities for sufficiently close binaries.
Although these simplified analytic studies can give much physical
insight into difficult questions of global fluid instabilities, 
fully numerical calculations remain essential for establishing
the stability limits of close binaries accurately and for following 
the nonlinear evolution of unstable systems all the way to complete 
coalescence. 

A number of different groups have now performed such calculations, using
a variety of numerical methods and focusing on different aspects of the
 problem. Nakamura and collaborators (see Nakamura \& Oohara 1998 and 
references therein)
were the first to perform 3D hydrodynamic calculations of binary 
NS coalescence, using a traditional Eulerian finite-difference code. 
Instead, RS used the 
Lagrangian method SPH (Smoothed Particle Hydrodynamics). They focused
on determining the ISCO for initial binary models in strict
hydrostatic equilibrium and calculating the emission of gravitational waves
from the coalescence of unstable binaries. Many of the results of RS were
later independently confirmed by New \& Tohline (1997) and Swesty,
Wang, \& Calder (1999), who used completely
different numerical methods but also focused on stability questions, and 
by Zhuge, Centrella, \& McMillan (1994, 1996), who also 
used SPH. Zhuge et al.\ (1996) also explored in detail the dependence of
the gravitational wave signals on the initial NS spins. 
Davies et al.\ (1994) and Ruffert et al.\ (1996, 1997) have
incorporated a treatment of the nuclear physics in their hydrodynamic
calculations (done using SPH and PPM codes, respectively), motivated
by cosmological models of GRBs.
All these calculations were performed in {\it Newtonian gravity\/}, with
some of the more recent studies adding an approximate treatment of
energy and angular momentum dissipation through the gravitational 
radiation reaction (e.g., Janka et al.\ 1999; Rosswog et al.\ 1999),
or even a full treatment of PN gravity to lowest order (Ayal et al.\ 2000;
Faber \& Rasio 2000).

All recent hydrodynamic calculations agree on
the basic qualitative picture that emerges for the final coalescence.
As the ISCO is approached, the secular orbital
decay driven by gravitational wave emission is dramatically accelerated
(see also LRS2, LRS3).
The two stars then plunge rapidly toward each other, and merge together 
into a single object in just a few rotation periods. In the corotating 
frame of the binary, the relative radial velocity of the two stars always 
remains very subsonic, so that the evolution is nearly adiabatic.
This is in sharp contrast to the case of a head-on collision between
two stars on a free-fall, radial orbit, where
shock heating is very important for the dynamics (RS1; Shapiro 1998).
Here the stars are constantly being held back by a (slowly receding)
centrifugal barrier, and the merging, although dynamical, is much more gentle. 
After typically $1-2$ orbital periods following first contact,
 the innermost cores of the 
two stars have merged and the system resembles a single, very elongated ellipsoid.
At this point a secondary instability occurs: {\it mass shedding\/} 
sets in rather abruptly. Material (typically $\sim10$\% of the total mass) 
is ejected through the outer Lagrange
points of the effective potential and spirals out rapidly.
In the final stage, the inner spiral arms widen and merge together, 
forming a nearly axisymmetric torus around the inner, maximally rotating
dense core. 

In GR, strong-field gravity between the masses in
a binary system is alone sufficient to drive a close circular 
orbit unstable. In close NS binaries, GR effects combine nonlinearly
with Newtonian tidal effects so that the ISCO is encountered
at larger binary separation and lower orbital frequency than 
predicted by Newtonian hydrodynamics alone, or GR alone for two point
masses. The combined effects
of relativity and hydrodynamics on the stability of close compact
binaries have only very recently begun to be studied,
using both analytic approximations
(basically, PN generalizations of LRS; see, e.g., 
Lai \& Wiseman 1997; Lombardi, Rasio, \& Shapiro 1997; 
Shibata \& Taniguchi 1997), as well as numerical 
calculations in 3D incorporating simplified treatments of 
relativistic effects 
(e.g., Baumgarte et al.\ 1998; Marronetti, Mathews \& Wilson 1998;  
Wang, Swesty, \& Calder 1998).
Several groups have been working on a fully general relativistic
calculation of the final coalescence, combining the techniques of 
numerical relativity and numerical hydrodynamics in 3D
(Baumgarte, Hughes, \& Shapiro 1999;
Landry \& Teukolsky 1999; 
Seidel 1998; Shibata \& Uryu 1999). 
However this work is still in its infancy, and only very preliminary results
of test calculations have been reported so far.

\section{Black Hole Formation}

The final fate of a NS--NS merger depends crucially on the NS EOS,
and on the extraction of angular momentum from the system during the 
final merger. For a stiff NS EOS, it is by no means
certain that the core of the final merged configuration will collapse
on a dynamical timescale to form a BH. One reason is that the Kerr
parameter $J/M^2$ of the core may exceed unity for extremely stiff
EOS (Baumgarte et al.\ 1998), although Newtonian and PN 
hydrodynamic calculations suggest that this is never the case
(see, e.g., Faber \& Rasio 2000). 
More importantly, the rapidly rotating core may in fact be 
dynamically stable. Take the obvious example of a system containing two 
identical $1.35\,M_\odot$ NS. The total baryonic mass of the system
for a stiff NS EOS is then about $3\,M_\odot$. Almost independent of 
the spins of the NS, all hydrodynamic calculations suggest that about
$10\%$ of this mass will be ejected into the outer torus, leaving at
the center a {\it maximally rotating\/} object with baryonic mass 
$\simeq2.7\,M_\odot$ (Any hydrodynamic merger process that leads to mass
shedding will produce a maximally rotating object since the system will
have ejected just enough mass and angular momentum to reach its new,
stable quasi-equilibrium state). Most stiff NS EOS (including the
recent ``AU'' and ``UU'' EOS of Wiringa et al.\ 1988) allow stable,
maximally rotating NS with baryonic masses exceeding
$3\,M_\odot$ (Cook, Shapiro, \& Teukolsky 1994), i.e., well above the mass
of the final merger core. Differential rotation (not taken into account in the
calculations of Cook et al.\ 1994) can further increase this maximum stable 
mass very significantly (see Baumgarte, Shapiro, \& Shibata 2000).
However, for slowly rotating stars, the same EOS give
maximum stable baryonic masses in the range $2.5-3\,M_\odot$.
Thus the final fate of the merger depends critically on its rotational
profile and total angular momentum. 

Note that other processes, such as 
electromagnetic radiation, neutrino emission, and the development of
various secular instabilities (e.g., r-modes), which may also lead to angular
momentum losses, take place on timescales much longer than the dynamical
timescale (see, e.g., Baumgarte \& Shapiro 1998, who show that
neutrino emission is probably negligible). These processes are
therefore decoupled from the hydrodynamics of the coalescence.
Unfortunately their
study is plagued by many fundamental uncertainties in the microphysics.

The question of the final fate of the merger also depends crucially
on the evolution of the fluid vorticity during the final coalescence.
Close NS binaries  are likely to be {\it nonsynchronized\/}. Indeed,
the tidal synchronization time 
is almost certainly much longer than the orbital decay
time (Kochanek 1992; Bildsten \& Cutler 1992).
For NS binaries that are far from synchronized,
the final coalescence involves
some new, complex hydrodynamic processes (Rasio \& Shapiro 1999).
Consider for example the case of an irrotational system (containing
two nonspinning stars at large separation; see LRS3).
Because the two stars appear to be counter-spinning in the corotating
frame of the binary, a {\it vortex sheet\/}  (where the tangential velocity
jumps discontinuously by $\Delta v\sim 0.1\,c$) appears when the stellar 
surfaces come into contact.
Such a vortex sheet is Kelvin-Helmholtz unstable on all 
wavelengths and the hydrodynamics is therefore  extremely
difficult to model accurately given the limited spatial
resolution of 3D calculations.
The breaking of the vortex sheet generates some turbulent
viscosity so that the final configuration may no longer be
irrotational. In numerical simulations, however, vorticity is
quickly generated through spurious shear viscosity, and the 
merger remnant is observed to evolve rapidly (in just a few
rotation periods) toward uniform rotation.

The final fate of the merger will be affected drastically by these
processes. In particular, the shear flow inside the merging stars 
(which supports a highly triaxial shape; see Rasio \& Shapiro 1999) may
in reality persist long enough to allow a large fraction of the total
angular momentum
in the system to be radiated away in gravitational waves. In this case
the final merged core may resemble a Dedekind ellipsoid, i.e., it will have
a triaxial shape supported entirely by internal fluid motions, but with
a stationary shape in the inertial frame (so that it no longer
radiates gravitational waves). 
This state will be reached on the gravitational radiation reaction
timescale, which is no more than a few tens of rotation periods.
On the (possibly much longer) {\it viscous timescale\/}, the core will
then evolve to a uniform, slowly rotating state and will likely collapse to
a BH.
In contrast, in all 3D numerical simulations performed to date,
the shear is quickly dissipated, so that gravitational radiation
never gets a chance to extract more than a small fraction ($\sim10$\%)
of the angular momentum, and the final core appears to be a uniform,
maximally rotating object exactly as in calculations starting
from synchronized binaries. However this behavior is most likely
an artefact of the large spurious shear viscosity present in the
3D simulations. 

In addition to their obvious significance for gravitational wave emission,
these issues are also of great importance for models 
of GRBs that depend on energy extraction from a torus of material around
the central BH. Indeed, if a large fraction of the total angular momentum
is removed by the gravitational waves, rotationally-induced mass shedding 
may not occur at all during the merger, leaving a BH with no surrounding 
matter and no way of extracting energy from the system.

\section*{Acknowledgements}

This work was supported by NSF Grant AST-9618116, NASA ATP Grant NAG5-8460, 
and by an Alfred P.\ Sloan Research Fellowship. The computational
work was also supported by the National Computational Science Alliance
and utilized the SGI/Cray Origin2000 supercomputer at NCSA.

%INDEX%%%%%%%%%%%%%%%%%%%%%%%%%%%%%%%%%%%%%%%%%%%%%%%%%%%%%%%%%%%%%%%
\clearpage
\addcontentsline{toc}{section}{Index}
\flushbottom
%\printindex
%%%%%%%%%%%%%%%%%%%%%%%%%%%%%%%%%%%%%%%%%%%%%%%%%%%%%%%%%%%%%%%%%%%%%

\end{document}